\documentclass[aps,prd,onecolumn,superscriptaddress]{revtex4}
\usepackage{graphicx}
\usepackage{amsmath}
\usepackage{epsfig, epstopdf}
\usepackage{amsmath,amsfonts,amssymb}

\newcommand{\nc}{\newcommand}

\usepackage{color}
\nc{\ba}{\begin{eqnarray}}
\nc{\ea}{\end{eqnarray}}

\nc{\x}{{\bf{x}}}
\nc{\bfk}{{\bf{k}}}

\begin{document}
\baselineskip=12pt
\def\black{\textcolor{black}}
\def\red{\textcolor{black}}
\def\blue{\textcolor{blue}}
\def\green{\textcolor{black}}
\def\be{\begin{equation}}
\def\ee{\end{equation}}
\def\bea{\begin{eqnarray}}
\def\eea{\end{eqnarray}}
\def\orc{\Omega_{r_c}}
\def\om{\Omega_{\text{m}}}
\def\E{{\rm e}}
\def\bearst{\begin{eqnarray*}}
\def\eearst{\end{eqnarray*}}
\def\peleven{\parbox{11cm}}
\def\peffec{\peight{\bearst\eearst}\hfill\peleven}
\def\pspace{\peight{\bearst\eearst}\hfill}
\def\ptwelve{\parbox{12cm}}
\def\peight{\parbox{8mm}}

\title{The Schwarzschild metric and the Friedmann equations from Newtonian Gravitational collapse}

\author{Eduardo I. Guendelman}
\email{guendel@bgumail.bgu.ac.il},  
\author{Arka Prabha Banik}
  \email{arkaprab@post.bgu.ac.il}
\author{Gilad Granit}
  \email{granitg@post.bgu.ac.il}
\author{Tomer Ygael}
  \email{tomeryg@post.bgu.ac.il}
 \affiliation{Ben Gurion University of the Negev, Department of Physics, Beer-Sheva, Israel}
\author{Christian Rohrhofer}
  \email{christian.rohrhofer@edu.uni-graz.at}
\affiliation{University of Graz, Institute of Physics, 8010 Graz,
Austria}

\date{\today}

\begin{abstract} 

As  is well known, the $0-0$ component of the Schwarzschild space can be obtained by the requirement that the geodesic of slowly moving particles match the Newtonian equation. Given this result, we show here that the remaining components can be obtained by requiring that the inside of a Newtonian ball of dust matched at a free falling radius with the external space determines that space to be Schwarzschild, if no pathologies exist. Also we are able to determine that the constant of integration that appears in the Newtonian Cosmology, coincides with the spatial curvature of the FLRW metric. These results are of interest at least in two respects, one from the point of view of its pedagogical value of teaching General Relativity without in fact using Einstein's equation and second, the fact that some results attributed
to General Relativity can be obtained without using General Relativity indicates that these results
are more general than the particular dynamics specified by General Relativity. 

\begin{center}
This essay has been written for the Gravity Research Foundation 2015 essay competition \\
\end{center}


\end{abstract}

\maketitle
\section{Introduction}
Many aspects of General Relativity can be derived without using Einstein’s Equations (EE). For example, it is well known that the 0-0 component of Schwarzschild metric can be derived from the requirement correspondence with the Newtonian limit. As shown in ref 1 , on which this essay is based, the  $g_{\bar{r}\bar{r}}$ component can be derived from matching the external space , assumed to be static,  to an internal dust shell. This derivation is very different from the derivations for  the  $g_{\bar{r}\bar{r}}$ component  explained in ref 2  and criticized in ref 3. We use the Newtonian Cosmology results, where a homogeneous and isotropic dust ball with co-moving dust particles can also shown to be studied.

\medskip

Finally, the self consistency  of the treatment produces the outcome that the constant of integration k can be interpreted as the spatial curvature in FRLW spacetime without using the EE .

\section{Newtonian Cosmology}

In a sphere of arbitrary size with homogeneous and isotropic in spacetime
will yield Friedmann's second equation without Cosmological constant. The proof is as follows:

\medskip

The equation of motion for a test mass $m$ located on the boundary of  a sphere shell be described in terms of a homogeneous positive parameter $R(t)$, where the coordinate of each particle expands according to $a(t)=constant \cdot R(t)$, where the constant depends on the particular particle therefore such equation  reads
\begin{equation}
\label{eq:a}
\ddot{a}=-\frac{G}{a^2} \left(\frac{4\pi}{3}a^3 \rho \right)= - \frac{4\pi G}{3} a\rho.
\end{equation}
which implies a similar equation for the universal expansion factor $R(t)$
\begin{equation}
\label{eq:newton1}
\ddot{R}=-\frac{G}{R^2} \left(\frac{4\pi}{3}R^3 \rho \right)= - \frac{4\pi G}{3} R\rho.
\end{equation}
Basically, this corresponds to Friedmann's second equation without a cosmological constant $\Lambda$ and zero pressure.
As the linear dimensions scale by $R(t)$, all co-moving volumes should scale by $R(t)^3$, that is a $1/R^{3}$ dependency for the density, which dilutes the matter as the sphere expands.

For deriving a second equation, we first consider mass conservation within comoving sphere,
\begin{equation}
\label{eq:energyconservation1}
\frac{d}{dt}\left(\frac{4\pi}{3}  R^{3}\rho  \right) = 0,
\end{equation}
where the internal mass $M$ inside the sphere should be constant . By performing the derivative and simplifying one $R$, the equation gets
\begin{equation*}
2R\dot{R}\rho + R\dot{R}\rho + R^2 \dot{\rho} = 0.
\end{equation*}
The second term can be eliminated by (\ref{eq:newton1}) and after restoring derivatives the equation
\begin{equation*}
\frac{d(\dot{R}^2)}{dt} = \frac{8\pi G}{3} \frac{d(\rho R^2)}{dt}
\end{equation*}
is obtained. Integration on both sides gives
\begin{equation*}
\dot{R}^2 = \frac{8\pi G}{3} \rho R^2 - \tilde{k},
\end{equation*}
and rewriting the arbitrary integration constant $\tilde{k}$ in a way to match the units $\tilde{k} \longrightarrow kc^2$ yields finally an equation, which corresponds to Friedmann's first equation in structure:  
\begin{equation}
\label{eq:fridman1}
\left(\frac{\dot{R}}{R}\right)^{2} = \frac{8\pi G}{3} \rho - k \left(\frac{c}{R}\right)^2
\end{equation}
The implication of the integration constant k is that it implies negative energy bound system, 
 positive energy system and zero energy free system for
 $k>0,k<0$ and $k=0$  respectively.
\section{Finding $g_{00}$}
As we all know the geodesic equation can be derived from the from
the principle of least action of the particle trajectory, the action
being the proper time along the trajectory of the particle in a certain
given metric. Now, that in order to obtain the Newtonian non relativistic limit, the correct Newtonian force equation coincides with the geodesic equation provided by following equations
\begin{eqnarray}
\frac{d^2 x}{dt^2} = - \nabla \phi \quad \textsl{where} \quad \phi = -\frac{GM}{r}
\end{eqnarray}
and
\begin{eqnarray}
g_{tt}=-\left(1-\frac{2GM}{r}\right)=-\left(1+2\phi\right)
\end{eqnarray}\\
\section{Finding $g_{\bar{r}\bar{r}}$}
The  $g_{\bar{r}\bar{r}}$ of the metric in the outside static region in our case,  has been the "elusive component". This component has not been calculated using matching to a Newtonian cosmology previously, here we will show that this is possible\cite{1}. In this section we will find $g_{\bar{r}\bar{r}}$ using the assumption that we have a co-moving observer satisfying $r = constant$. We will also assume that inside and at the boundary of the dust ball the radius evolves as $\bar{r}= r R(t)$, where $R(t)$ is determined by (\ref{eq:fridman1}). We also assume that for $\bar{r} > r R(t)$ the motion is on a radial geodesic and the metric is of the form
\begin{eqnarray}
ds^2 &=& - \left( 1 - \frac{2GM}{\bar{r}}\right) d\bar{t}^2 + A(\bar{r}) d\bar{r}^2 + \bar{r}^2 d \Omega^2 \\
d \Omega^2 &=& d \theta^2 + \sin^2 \theta d \phi^2
\end{eqnarray}
we labelled the time $\bar{t}$ because it may not be the same coordinate as the $t$ in the dust ball and the  $g_{tt}$ component was found in the previous section. Notice that we assume that the external metric is time independent.\\
A radially falling geodesic, meaning that $\theta = const$ and $\phi = const$, is fully described by the conservation of energy that results from that the metric outside is assumed to be static.
The geodesics are derived from the action 
\begin{eqnarray}
S = \int d\sigma \sqrt{-\frac{d x^\mu}{d \sigma} \frac{d x^\nu}{d \sigma} g_{\mu \nu}(x)}
\end{eqnarray}
The equation with respect to $\bar{t}$ is
\begin{equation}
\frac{d}{d\sigma}\left( \frac{\partial L}{\partial \dot{\bar{t}}}\right) = 0 \quad \textsl{where} \quad \dot{\bar{t}} = \frac{d \bar{t}}{d \sigma}
\end{equation}

This gives us
\begin{equation}
\gamma = \frac{\partial L}{\partial \dot{\bar{t}}} = \left( 1 - \frac{2GM}{\bar{r}} \right) \frac{d \bar{t}}{d \tau}
\end{equation}
where $\gamma$ is constant and $d\tau$ is the proper time.\\
Notice that since the spatial coordinates in the space $d x^i = 0$ we get 
\begin{equation}
d\tau = d t
\end{equation}
giving us
\begin{eqnarray}
d \tau^2 &=&  \left( 1 - \frac{2GM}{\bar{r}} \right) d \bar{t}^2 - A(\bar{r})d\bar{r}^2\\
\left( \frac{d \tau}{d t} \right)^2 = 1 &=& \left( 1 - \frac{2GM}{\bar{r}} \right) \left( \frac{d \bar{t}}{dt} \right)^2 - A(\bar{r}) \left( \frac{d\bar{r}}{dt} \right)^2
\end{eqnarray}
Using $(30)$ and $(31)$ we obtain
\begin{eqnarray}
\left( 1- \frac{2GM}{\bar{r}}\right)^{-1}\gamma^2 - A(\bar{r}) \left( \frac{d\bar{r}}{dt} \right)^2 = 1 
\end{eqnarray}
As we have seen, the consistency of the matching of the two spaces requires
 $\bar{r} = R(t) r$, furthermore, we assume that even the boundary of the dust shell
 free falls according to a co-moving observer, which means that the FLRW coordinate
 $r= constant$ and this allow then to solve for  $A(\bar{r})$, 
\begin{eqnarray}
A(\bar{r}) = - \left( 1 - \frac{\gamma^2}{\left( 1 - \frac{2GM}{\bar{r}} \right)}\right)\frac{1}{r^2 \left( \frac{d R}{dt}\right)^2}
\end{eqnarray}
We now use $(3)$ and get

\begin{eqnarray}
A(\bar{r}) = -\left( 1 - \frac{\gamma^2}{\left( 1 - \frac{2GM}{\bar{r}} \right)}\right)\frac{1}{r^2 k \left( \frac{1}{R} -1 \right)}
\end{eqnarray}
Simplifying this and expressing in terms of $\bar{r}$ we get
\begin{eqnarray}
A(\bar{r}) = \frac{1}{r^2}\frac{\gamma^2 - 1 + \frac{2GM}{\bar{r}}}{1 - \frac{2GM}{\bar{r}}}\frac{1}{k \left( -1 + \frac{r}{\bar{r}}\right)}
\end{eqnarray}

If we take the limit $\bar{r} \rightarrow \infty $, we see that 
$A(\bar{r}) \rightarrow -(\gamma^2 - 1)/k r^2 $. Asymptotic flatness would require 
$A(\bar{r}) \rightarrow 1$. 

The metric component $A(r)$ is free of singularities (real singularities, not coordinate singularities) and preserves its signature (one time and three spaces) and is asymptotically flat only if
\begin{eqnarray}
\label{A}
\gamma^2 - 1 = - k r^2 \quad \textsl{and} \quad \frac{2GM}{\bar{r}} = \frac{r^2 k}{R} = \frac{k r^3}{\bar{r}} \Rightarrow k = \frac{2GM}{r^3}
\end{eqnarray}
Notice that above the condition,(\ref{A}) $ k = \frac{2GM}{r^3}$ , when combined with the value of k, as given by
(\ref{the value of k}), $k = \frac{8 \pi G}{3} \rho(0)$ tells us the eminently reasonable relation,
\begin{eqnarray}
M=\frac{4}{3}\pi\rho_{0}r^{3}
\end{eqnarray}

which indicates that this choice has a good physical basis. Finally, all of this gives us
\begin{eqnarray}
A(\bar{r}) = \frac{1}{1 - \frac{2GM}{\bar{r}}}
\end{eqnarray}
Reproducing the Schwarzschild spacetime.

\section{Finding the geometric interpretation of the Newtonian Cosmology}
The main aim of this section is to show that for the geometric parameter defining the FRW space coincides with the Newtonian Energy in $k$ found from the interpretation of the Newtonian cosmology and to demonstrate that $k=\kappa$.

We start with a 3-sphere by considering first an embedding four dimensional Euclidean Space with metric
to  finally obtain the Infinitesimal line element in
FRW space as, 
\begin{eqnarray}
ds^{2}=-dt^{2}+R(t)^{2}\left(\frac{dr^{2}}{1-\kappa r^{2}}+r^{2}d\Omega^{2}\right)
\end{eqnarray}
We have assume  that that we have a co-moving observer which satisfies $r = const$. Independently of that,
in the FLRW space we can use everywhere (not just at the boundary) the barred radius  $\bar{r}=R(t)r$, which means $r=\frac{\bar{r}}{R(t)}$.
we will find $g_{\bar{r}\bar{r}}$ by the above-said transformation to get a relation between $g_{\bar{r}\bar{r}}$ with $\kappa$ .Matching this with the same value of $g_{\bar{r}\bar{r}}$ from the previous section, we can infer that $k=\kappa$.

A final consistency check is obtained now with that we have derived that the internal space is a Robertson Walker space  satisfying the standard Friedmann equations, with k being interpreted as the spatial curvature. Under this condition, we know the matching of this internal 
cosmology to Schwarzschild is consistent, as the analysis of the Oppenheimer- Snyder collapse model shows\cite{Weinberg}.
Here we have gone about this problem in the opposite way, showing that the matching of these two 
spaces imposes severe constraints that allows us to derive Schwarzschild space in the outside
and determine that the Newtonian constant of integration $k$ has  to be the spatial curvature of the internal FLRW internal space, all of this without using Einstein's equations .

\section{Conclusions}
In this essay, it has been found that matching a dust ball, whose dynamics
is governed by the Newtonian cosmology equations, containing a constant
of integration $k$ coincides with the geometrical parameter $\kappa$
that appears in FLRW used in the Newtonian Cosmology. These results
are of interest at least in two respects, one from the point of view
of its pedagogical value of teaching General Relativity without in
fact using Einstein's equation and second, the fact that some results
attributed to General Relativity can be obtained without using General
Relativity indicates that these results are more general than the
particular dynamics specified by General Relativity\citep{5}. If we ponder minutely, the paper \cite{ref5}  goes very far just using Newtonian physics, but is missing a space-time interpretation of the solutions of Newton’s laws for gravitationally interacting particles. We might try to find them properly even in presence of pressure in our next venture.
 Finally, an interesting question arises that whether our derivation holds only in the weak field approximation or not. Notice that indeed in parts of our arguments we have used
the weak field approximation, like when we derived the $0-0$ component
of the metric, then on this basis, we derived the $r-r$ component
by a process of matching to the internal collapsing ball of dust.
But if we take the point of view that we trust the metric of the collapsing
dust beyond the weak field approximation, the situation will be different,
in this case our derivation could have validity beyond the weak field
approximation, this is a question to be studied. One should also point
out that in GR, as far as the post-Newtonian approximation is concerned,
the corrections to the the $0-0$ component of the metric appear at
the same other as the first corrections to the $r-r$ components,
so it is in a sense puzzling that the $r-r$ component has been more
elusive to find by a simple derivation, as compared to the case of
the $0-0$ component.

\end{document}